\pdfoutput=1

\documentclass[11pt]{article}

\usepackage[final]{acl}

\usepackage{times}
\usepackage{latexsym}

\usepackage[T1]{fontenc}

\usepackage[utf8]{inputenc}

\usepackage{microtype}
\usepackage{float}

\usepackage{inconsolata}

\usepackage{graphicx}

\usepackage{amsmath}
\usepackage{booktabs}
\usepackage{multirow}
\usepackage{pifont}
\usepackage[inline]{enumitem}
\usepackage{array}
\usepackage{paralist}
\usepackage[font=small,labelfont=bf]{caption}
\usepackage{subcaption}   
\usepackage{changepage}
\usepackage{xcolor,colortbl} 
\usepackage{pgfplots}
\usepackage{CJKutf8}
\usepackage{placeins}

\pgfplotsset{compat=1.18}

\definecolor{CBBlue}{HTML}{0072B2}
\definecolor{CBOrange}{HTML}{D55E00} 
\newcommand{\cmark}{\textcolor{CBBlue}{\ding{51}}} 
\newcommand{\xmark}{\textcolor{CBOrange}{\ding{55}}}

\newcolumntype{L}[1]{>{\raggedright\arraybackslash}p{#1}}
\newcolumntype{C}[1]{>{\centering\arraybackslash}p{#1}}
\newcolumntype{B}[1]{>{\raggedright\arraybackslash\bfseries\normalsize}p{#1}}
\newcolumntype{S}[1]{>{\centering\arraybackslash}m{#1}}

\renewcommand{\arraystretch}{1.1} 

\title{SCENEBench: An Audio Understanding Benchmark Grounded in Assistive and Industrial Use Cases}

\author{
\textbf{Laya Iyer}$^{1}$ \quad
\textbf{Angelina Wang}$^{2}$ \quad
\textbf{Sanmi Koyejo}$^{1}$ \\
$^{1}$Stanford University \quad
$^{2}$Cornell Tech \\
\texttt{laya, sanmi@stanford.edu} \quad
\texttt{angelina.wang@cornell.edu}
}

\begin{document}
\maketitle
\begin{abstract}
Advances in large language models (LLMs) have enabled significant capabilities in audio processing, resulting in state-of-the-art models now known as Large Audio Language Models (LALMs). However, minimal work has been done to measure audio understanding beyond automatic speech recognition (ASR). This paper closes that gap by proposing a benchmark suite, SCENEBench (Spatial, Cross-lingual, Environmental, Non-speech Evaluation), that targets a broad form of audio comprehension across four real-world categories: \texttt{background} \texttt{sound} \texttt{understanding}, \texttt{noise localization}, \texttt{cross-linguistic} \texttt{speech} \texttt{understanding}, and \texttt{vocal characterizer recognition}. These four categories are selected based on understudied needs from accessibility technology and industrial noise monitoring. In addition to performance, we also measure model latency. The purpose of this benchmark suite is to assess the audio beyond just what words are said --- rather, in \textit{how} they are said and the non‑speech components of the audio. Because our audio samples are synthetically constructed (e.g., by overlaying two natural audio samples), we further validate our benchmark against 20 natural audio items per task, sub-sampled from existing datasets to match our task criteria, to assess ecological validity. We assess five state-of-the-art LALMs and find critical gaps: the performance in each task is varied, with some tasks having performance below random chance and others with high accuracy. These results provide direction for targeted improvements in model capabilities.
\end{abstract}

\section{Introduction}

Remarkable strides in natural language understanding have powered a wide range of applications in search, conversation, and information retrieval. As the capabilities of these models improve, we must develop methods to evaluate them effectively. 

Speech, a widely shared mode of human communication, is an extension for text-based large language models. However, understanding spoken language is not limited to transcription. Proper audio comprehension involves recognizing tone, emotion, background noise, environmental context, speaker intent, and more. These elements often coexist and can significantly impact meaning. They also enable practical systems, such as assistive devices or captioning tools that describe traffic sounds and approaching sirens for individuals with hearing impairments, as well as telehealth and well-being systems that detect coughs, or sobs in patient speech.

Companies releasing LALMs, such as GPT-4o \cite{openai2024gpt4o} and Qwen2-Audio \cite{chu2024qwen2audio}, advertise capabilities beyond ASR—for example, Alibaba states that Qwen2-Audio “can transcribe speech and identify audio info … including spoken words, music and ambient noises” \cite{AlibabaQwen2AudioBlog2024}. However, current evaluation strategies primarily measure speech recognition, not audio understanding. That is, they \textit{assess what words were said, not how they were said, or the non-speech components of the audio.}

In this paper, we present a new benchmark suite, SCENEBench, for evaluating a broader conception of audio understanding in LALMs, grounded in the needs of accessibility technology and industrial noise monitoring. Many prior benchmarks focus on controlled, single-modality, or clean scenarios. Our benchmark tests four categories that reflect real-world complexity: \texttt{background} \texttt{sound} \texttt{understanding}, \texttt{noise localization, cross-linguistic speech understanding, and vocal characterizers} (e.g., crying). All of these are measured along with model latency, a parallel dimension of our benchmark suite.

\section{Related Works}

In \hyperref[sec:benchmarks]{Section~\ref*{sec:benchmarks}}, we review and break down the scope of prior benchmarks for audio and speech understanding \hyperref[tab:benchmark_comparison]{(Table~\ref*{tab:benchmark_comparison})}. In \hyperref[sec:gaps]{Section~\ref*{sec:gaps}}, we analyze gaps through two high-impact settings with clear stakes: accessibility and industrial monitoring. We will also discuss how our benchmark suite generalizes to other contexts.

\begin{table*}[ht]
\centering
\setlength{\tabcolsep}{3pt}
\renewcommand{\arraystretch}{1.05}
\footnotesize

\begin{tabular}{@{}L{2.3cm} C{1.4cm} C{0.95cm} L{2.1cm} S{1.5cm} S{1.1cm} C{0.95cm} C{0.90cm} C{0.95cm} C{0.95cm} C{0.95cm}@{}}
\toprule
\textbf{Benchmark} & \textbf{Type} & \textbf{\shortstack{Multi\\turn}} & \textbf{Primary Focus} & \textbf{Clips ($\approx$)} & \textbf{\#Tasks} &
\textbf{\shortstack{\texttt{Bkgd}\\\texttt{Sound}}} & \textbf{\shortstack{\texttt{Noise}\\ \texttt{Loc.}}} & \textbf{\shortstack{\texttt{Cross}\\\texttt{Ling.}}} & \textbf{\shortstack{\texttt{Vocal}\\\texttt{Chars}}} & \textbf{Latency} \\
\midrule
\textbf{SCENEBench} & MC+FRQ & \cmark & Audio \& speech understanding & 16k & 4 & \cmark & \cmark & \cmark & \cmark & \cmark \\
\textbf{AudioBench}~\cite{AudioBench2024} & MC+FRQ & \xmark & ASR, scene \& voice & 100k & 8 & \cmark & \xmark & \xmark & \xmark & \xmark \\
\textbf{MMAU}~\cite{MMAU2024} & MC & \xmark & Multi\mbox{-}task reasoning & 10k & 27 & \cmark & \xmark & \xmark & \xmark & \xmark \\
\textbf{AIR\mbox{-}Bench}~\cite{AIRBench2024} & MC+FRQ & \cmark & Generative comprehension & 21k & 20 & \xmark & \xmark & \xmark & \xmark & \xmark \\
\textbf{MARBLE}~\cite{MARBLE2023} & MC & \xmark & Music classification & 25.9k & 18 & \xmark & \xmark & \xmark & \xmark & \xmark \\
\textbf{Clotho\mbox{-}AQA}~\cite{ClothoAQA2022} & FRQ & \xmark & Environmental QA & 1.9k & 1 & \cmark & \xmark & \xmark & \xmark & \xmark \\
\textbf{SoundCheck}~\cite{SoundCheck2023} & Audit & \xmark & Dataset quality audit & 3M & 7 & \cmark & \xmark & \xmark & \xmark & \xmark \\
\textbf{SONAR}~\cite{SONAR2023} & CLS & \xmark & Deepfake detection & 2.2k & 1 & \xmark & \xmark & \xmark & \xmark & \xmark \\
\textbf{CAVA}~\cite{CAVA2025} & Task & \cmark & Voice\mbox{-}assistant behaviour & 6.4k & 6 & \xmark & \xmark & \xmark & \xmark & \cmark \\
\textbf{AU-Harness} ~\cite{surapaneni2025auharnessopensourcetoolkitholistic} & MC & \xmark & ASR, scene \& voice & 5k & 6 & \cmark & \cmark & \xmark & \xmark & \xmark \\
\textbf{LMMs-Eval} ~\cite{Zhang_2025} & MC & \xmark & Multimodal eval & 2k & 8 & \cmark & \xmark & \cmark & \xmark & \xmark \\
\bottomrule
\end{tabular}

\caption{\textbf{Benchmark comparison with standardized size and coverage.}
\emph{Type}: MC = multiple choice, FRQ = free response, CLS = classification, Audit = dataset audit.
\emph{Clips ($\approx$)} reports total items when stated or derivable; \emph{\#Tasks} counts distinct task families (— where not reported). Coverage columns indicate whether each suite evaluates that dimension.}
\label{tab:benchmark_comparison}
\end{table*}

\subsection{Benchmarks for Audio and Speech Understanding}
\label{sec:benchmarks}

Several benchmarks evaluate model capabilities on speech and audio inputs, each making design choices to hone in on their main focus.

\textbf{AudioBench} \citep{AudioBench2024} prioritizes broad coverage and automatically gradable tasks. This design enables reproducible cross-model comparisons at a large scale and lowers annotation cost. However, it deemphasizes phenomena that are harder to capture and grade automatically. \textbf{MMAU} \citep{MMAU2024} adopts a multiple choice format across diverse tasks to improve inter-annotator agreement, reduce evaluation variance and prompt sensitivity, and simplify scoring across large model suites. The MC constraint makes results comparable and stable, though it limits models’ ability to demonstrate open-ended reasoning, justifications, or descriptions of subtle acoustic attributes. \textbf{AIR-Bench} \citep{AIRBench2024} introduces open-ended audio Qs, but places a large emphasis on music and environmental sounds rather than detailed speech pragmatics and paralinguistics. \textbf{CAVA} \citep{CAVA2025} targets voice-assistant behavior (e.g., instruction following, latency), aligning with real deployment concerns. That focus surfaces agentic performance and responsiveness under constraints, while probing less of the fine-grained acoustic understanding. Other efforts, such as \textbf{SoundCheck} \citep{SoundCheck2023} and \textbf{SONAR} \citep{SONAR2023}, audit datasets or test deepfakes, serving specific goals rather than audio understanding as a whole.

Our benchmark suite is designed to complement the space of existing audio benchmarks by prioritizing the following: adding targeted evaluations for our four tasks and including free-response scoring where it is required. These are areas that are underrepresented in the prioritization of existing benchmarks. A side-by-side comparison with prior benchmarks appears in \hyperref[tab:benchmark_comparison]{Table~\ref*{tab:benchmark_comparison}}. These gaps matter most in high-stakes, real-world deployments and we will discuss them further in the next section.

\subsection{Use-Case Driven Gaps}
\label{sec:gaps}
From a use-case standpoint, existing audio benchmarks are still biased toward “clean-room” speech-recognition or captioning scenarios and rarely touch the two domains where errors are most consequential:  
\textbf{(i) accessibility} and \textbf{(ii) industrial sound monitoring}.  

\paragraph{Accessibility} Everyday scenes feature cross-talk, traffic, devices, and emotional or whispered speech. For Deaf or hard-of-hearing (DHH) users, useful support goes beyond transcription to sound awareness \citep{Jain2019CHI_SoundAwareness,Bragg2016ASSETS_PersonalizableDetector}. Some examples of crucial instances requiring the surfacing of salient non-speech events include hearing for sirens and their state (approaching vs.\ receding), and disambiguating paralinguistic cues (fatigue, sobs, whispers) \citep{Findlater2019Sound,SoundNarratives2024,VisibleNuances2023}. 

Wearable and IoT systems that detect sirens and vocalizations and relay haptic/onscreen alerts have been shown to reduce risk \cite{Chin2023,Salem2023}. Conversely, background noise (cross-talk, broadband noise) can degrade ASR rather than enrich it, underscoring the need for models to understand situational context, not just foreground words \cite{AIMultiple2025}.

\paragraph{Industrial monitoring}
Factory floors and labs demand early detection of anomalous machine sounds that often occur under speech or ambient noise. Public datasets such as ToyADMOS and MIMII established common testbeds for acoustic anomaly detection (AAD) in machine condition monitoring \cite{koizumi2019toyadmos,purohit2019mimii}. More recently, MIMII-DG introduced domain shifts across machine types and recording setups to test generalization of these models which is often the key blocker in deployment \cite{dohi2022mimiidg}. Missing or misclassifying subtle cues translates directly into safety risks.

\paragraph{Other applications}
Similar requirements recur in telehealth and well-being such as, cough and respiratory-symptom screening, fatigue/sleepiness cues \citep{orlandic2021coughvid,sharma2020coswara,schuller2011compareSleepiness}, in smart-home/IoT for domestic activity and rare-event detection such as, glass breaks \citep{mesaros2018dcaseDomestic,mesaros2017dcaseRare}, transportation and public safety, such as siren/horn awareness in urban scenes \citep{gemmeke2017audioset}, and AR/VR and robotics, where spatialized audio cues support navigation and interaction \citep{gordon2020soundspaces,chen2020audiogoal}.

\paragraph{Rationale for our benchmark.}
Across these settings, four failure modes appear repeatedly: important events may occur in the background; spatial/proximity cues are lost in audio understanding; transcription quality drops under multilingual spans; and paralinguistic information carried by non-speech is underspecified. SCENEBench turns each failure mode into a targeted evaluation. These four tasks are necessary because they probe salient audio-understanding capabilities tied to our use cases, and they are motivated by concrete failure patterns and potential user benefit. We exclude full spatial-audio setups and long-form clips in this release to keep the benchmark minimal, reproducible, and aligned with real-time assistive and monitoring scenarios. Taken together, these task choices yield a suite that diagnoses where current LALMs succeed and where they still fail.

\paragraph{Rationale for our tasks}
We did not choose our four evaluation dimensions arbitrarily. In reviewing the literature described in the previous two sections, the same failure modes appeared again and again: systems that ignore salient background sounds (sirens, alarms, machine tones), fail to track motion or distance cues under noise, normalize multilingual or code-switched speech to a single language, and miss non-speech vocal events such as coughs, laughs, or whispers that matter for situational awareness. SCENEBench directly instantiates these documented issues as four tasks—background sound understanding, noise localization, cross-linguistic speech understanding, and vocal characterizers—so that we can systematically measure where current LALMs break. We view this suite as diagnostic rather than exhaustive: it targets the most frequently reported, high-impact gaps in the literature, while leaving room for future extensions.

\section{Methods}

\label{sec:methods}
Our benchmark suite\footnote{Data and code: \href{https://github.com/layaiyer1/SCENEbench}{https://github.com/layaiyer1/SCENEbench}.} is constructed on top of existing datasets, to which we apply task-specific transformations. In this section, we describe the data construction process for each of the four tasks, highlighting how it diverges from conventional uses of the source dataset. We evaluate five leading models including GPT-4o \cite{openai2024gpt4o}, Qwen2-Audio-7B-Instruct \cite{chu2024qwen2audio}, and Gemini-2.5 \cite{comanici2025gemini25} to reveal the current limitations of LALMs in audio-first contexts.

\subsection{Tasks}
Our benchmark comprises four task types designed to probe distinct dimensions of audio understanding in real-world settings. In \hyperref[sec:ambient]{Section~\ref*{sec:ambient}}, we test \texttt{background sound understanding} by embedding environmental audio under speech and assessing models' ability to identify it. \hyperref[sec:distance]{Section~\ref*{sec:distance}} introduces a \texttt{noise localization} task based on amplitude change, targeting scenarios like siren detection for accessibility.  \hyperref[sec:crosslingual]{Section~\ref*{sec:crosslingual}} examines \texttt{cross-linguistic} robustness by asking to transcribe utterances in multiple languages. Finally, \hyperref[sec:vocal]{Section~\ref*{sec:vocal}} focuses on paralinguistic cues by testing whether models can recognize and label \texttt{vocal characterizers} such as laughter and whispering. In \hyperref[sec:latency]{Section~\ref*{sec:latency}}, we describe a parallel dimension of our benchmark, model latency, across all our tasks to measure real-time performance constraints.

\subsubsection{Background Sound Understanding}
\label{sec:ambient}

The first task examines whether a model can identify background noise layered under speech. While prior works such as WSJ0-2mix \citep{hershey2016deepclustering,isik2016deepclustering} and Libri2Mix (LibriMix) \citep{cosentino2020librimix} target speaker separation, few benchmarks probe environmental sounds under speech. We therefore overlay ESC-50 categories \citep{Piczak2015ESC50} onto DailyTalk utterances \citep{lee2023dailytalk}, using two voices (higher vs.\ lower pitch) to create two versions of each of 2{,}000 clips. The background noise and speech are overlaid with original volumes, which we discuss the limitations of in \hyperref[sec:limitations]{Section \ref{sec:limitations}}. Scoring is hierarchical: We evaluate the background-noise task with three prompts: (FR1) free-response description of all audible sounds; (FR2) targeted follow-up that explicitly asks for the background sound (issued only if FR1 omitted it); and (MC1) 4-way forced choice of the background category. Scoring is hierarchical: FR1 (describe audio.) is correct if the free-response mentions any background noise category; FR2 (describe background noise.) is credited if the correct ESC-50 class is named either in FR1 (in which case FR2 receives full credit without being asked) or in the FR2 follow-up; the 4-way multiple-choice probe is administered for all clips and scored independently. Exact prompt wordings appear in \hyperref[app:ambient-prompts]{Appendix~\ref*{app:ambient-prompts}}. Outputs are cleaned (lowercasing, punctuation stripping) and matched to a per-class synonym list with simple negation/uncertainty rejection (e.g., “no siren,” “not a dog,” “unsure”). This three-tier design deliberately separates spontaneous salience (FR1) from targeted retrieval (FR2) and discriminability (MC1), revealing whether failures arise from omission, misnaming, or confusion among plausible classes.

\subsubsection{Noise Localization}
\label{sec:distance}

This task evaluates models' ability to detect dynamic volume patterns in audio, simulating spatial motion through amplitude modulation. We created a dataset from the ESC-50 environmental sound corpus by applying three distinct volume envelopes to 2,000 source audio clips, yielding 6,000 total samples. Each original sound underwent three transformations: (1) approaching sound source, where amplitude scales from 10\% to 100\% over the clip duration;
(2) receding source, with amplitude linearly scaling from 100\% to 10\%; and
(3) oscillating movement past the listener, where amplitude follows a sinusoidal pattern (4 complete cycles) between 20\% and 100\%.

Models are evaluated using two complementary prompts. First (FR1), a general description prompt asks models to describe all auditory characteristics. Second (FR2), a follow-up position prompt specifically queries about spatial relationships and movement patterns relative to the sound source. Exact prompt wordings appear in \hyperref[app:noise-local-prompt]{Appendix~\ref*{app:noise-local-prompt}}. Responses are automatically scored as correct if they mention the appropriate motion pattern or appropriate synonyms (e.g., "approaching," "moving away," "oscillating").

\subsubsection{Cross-Linguistic Sound Recognition}
\label{sec:crosslingual}
We evaluate multilingual span transcription by transforming \textsc{DailyTalk} transcripts into controlled language-mixed stimuli: for each longest turn (2{,}541 total), we translate contiguous spans (\(\approx\)30\%) into one of four languages (Mandarin, Spanish, Hindi, Portuguese) via Google Translate, retain only items passing back-translation similarity \(>\!0.9\), and synthesize audio with ElevenLabs multilingual TTS. The resulting audio files, containing multilingual sentences (e.g., ``I have a fifteen-day vacation \begin{CJK*}{UTF8}{gbsn}我想拥有一个\end{CJK*} trip to England''), were presented to various LALMs for transcription evaluation (FR1), with performance measured by similarity between model transcriptions and the reference multilingual sentences. Because high-quality human recordings with natural code-switching across these languages are scarce, we adopt this synthetic route for coverage and control; we note its limits briefly here and discuss them in \hyperref[sec:limitations]{Section \ref{sec:limitations}}. As a result, this task, multilingual span transcription is a proxy behavior for code-switching.

\subsubsection{Vocal Characterizers}
\label{sec:vocal}

We target non-speech vocal traits—cough, cry, laugh, sneeze, yawn, mumble, and whisper—that carry communicative cues without requiring emotion inference. We deliberately avoid direct emotion classification due to documented ethical concerns about reductive labeling and potential harm \citep{stark2021ethics}. Our evaluation instead asks models first to briefly describe each clip (FR1), then to perform a 7-way classification over the vocal categories (MC1). The dataset aggregates publicly available repositories: \textsc{Nonspeech7k} for cough/cry/laugh/sneeze/yawn \citep{nonspeech7k}, \textsc{CapSpeech-AgentDB-Audio} for mumble/whisper \citep{capspeechagentdb}, with additional mumble from \textsc{vocal\_bursts\_taxonomy\_100\_clean\_wds} \citep{vocalbursts100} and whisper from \textsc{asmr} \citep{asmrrepo}. The final set contains 4{,}006 clips across five reported labels (632 cough, 1{,}791 cry, 1{,}133 laugh, 236 sneeze, 214 yawn), plus mumble and whisper for the 7-way classification.

\begin{figure*}[!t]
  \centering
  \includegraphics[width=\linewidth]{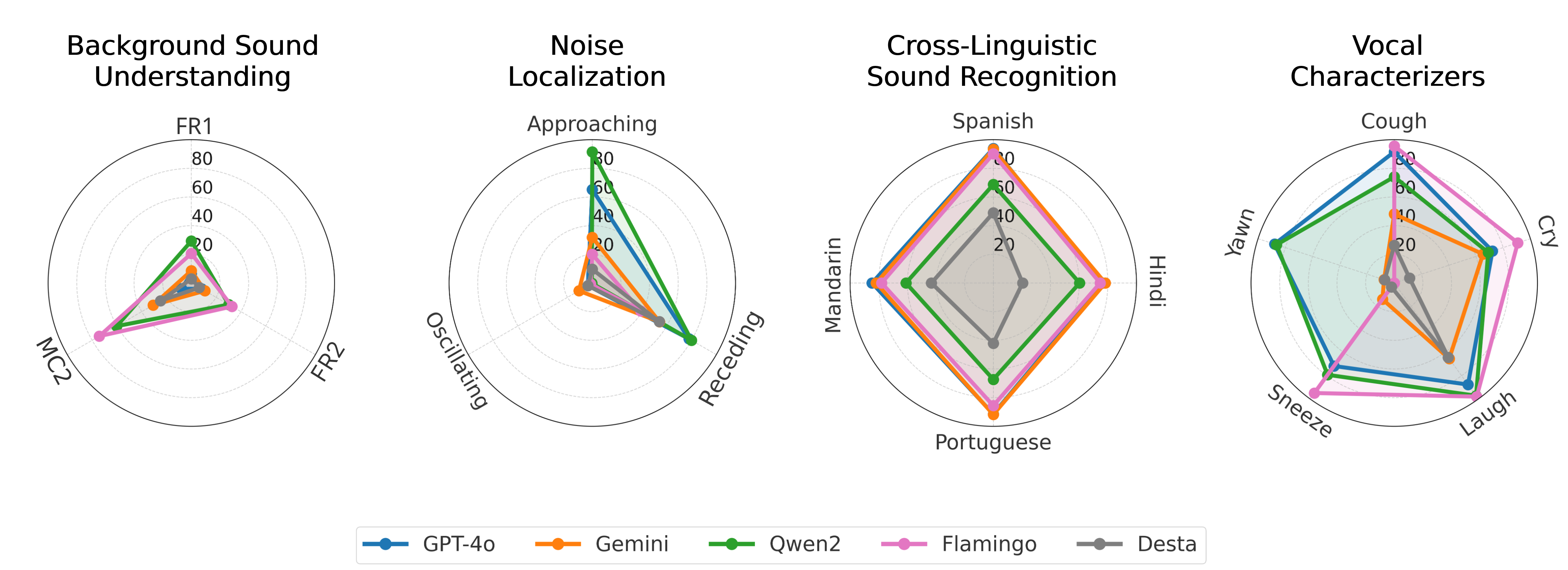}
  \caption{\textbf{Summary radar charts across tasks.} Each axis is a task-specific category; values are percentages (or mean similarity\,\%). Legend shown once: GPT-4o (blue), Gemini (orange), Qwen2 (green), Flamingo (pink), Desta (gray).}
  \label{fig:summary_spiders}
\end{figure*}

\subsection{Latency as a Dimension}
\label{sec:latency}

Beyond accuracy, we report latency for local models only; cloud/API models are excluded from timing comparisons. For each model invocation (one prompt–audio query), we record a the duration
between when the call is made and when the complete textual response is returned to our harness. We summarize latency across all four tasks using the median and interquartile range, and we report per-task medians to show how latency varies with input content and prompt type.

\subsection{Models Evaluated}

We benchmark five state-of-the-art LALMs models with audio capabilities: GPT-4o (OpenAI, USA) \cite{openai2024gpt4o}, Gemini 1.5 (Google DeepMind, USA/UK) \cite{comanici2025gemini25}, Qwen2-Audio (Alibaba DAMO Academy, China) \cite{chu2024qwen2audio}, Audio-Flamingo-3 (NVIDIA, USA) \cite{goel2025audioflamingo3advancing}, and DeSTA2-8B-beta (National Taiwan University + NVIDIA, Taiwan) \cite{lu2024developing}. These models span a range of architectures, training paradigms, and geographic origins.

We selected these five models to span the design space that is most relevant to our tasks and to balance fairness with reproducibility: they span commercial models (GPT-4o, Gemini) and open-weights models (Audio-Flamingo-3, Qwen2-Audio-7B, DeSTA2-8B-beta); and they are recent, widely used baselines that claim multilingual and non-speech capability aligned with our tasks. We exclude speech-only ASR and music-specialized models, in this paper, because they cannot run the full suite without additional components that would confound comparisons.

\section{Results}


We report results across four task types (summary in \hyperref[fig:summary_spiders]{Figure~\ref*{fig:summary_spiders}}), each designed to probe a distinct dimension of audio understanding. In \hyperref[sec:ambient-results]{Section~\ref*{sec:ambient-results}}, we analyze models’ ability to detect background sounds layered under speech. \hyperref[sec:distance-results]{Section~\ref*{sec:distance-results}} evaluates how well models can estimate the direction of background noise. \hyperref[sec:crosslingual-results]{Section~\ref*{sec:crosslingual-results}} presents findings on transcription accuracy in multilingual contexts. Finally, \hyperref[sec:vocal-results]{Section~\ref*{sec:vocal-results}} examines recognition of non-speech vocalizations, such as laughs and coughs, to assess affective and paralinguistic sound understanding.

\subsection{Background Sound Understanding}
\label{sec:ambient-results}

Each clip contains foreground speech with an ESC\textendash50 background sound. Models first give a free response (FR1: “describe the audio.”). If FR1 fails to mention the background, we ask a specified follow\textendash up (FR2: “name any specific background sound.”). Finally, the model answers a four\textendash way multiple\textendash choice question (MC1). For scoring, if the model already got FR1 right, we count FR2 as correct even when FR2 was not asked (so FR2 is counted for all clips). “Unsure/Cannot tell” is marked incorrect. Bars show means with 95\% CIs; horizontal dotted line mark chance (25\% for MC1). For readability, we report key outcomes here; full statistical details appear in \hyperref[app:stats]{Appendix~\ref*{app:stats}}. The 95\% CIs are extremely tight ($\leq$\,$\sim$1.5\,pp half–width at $N{=}4000$), so they are not visually distinguishable on the bars. For completeness, they are included in \hyperref[app:stats]{Appendix~\ref*{app:stats}}.

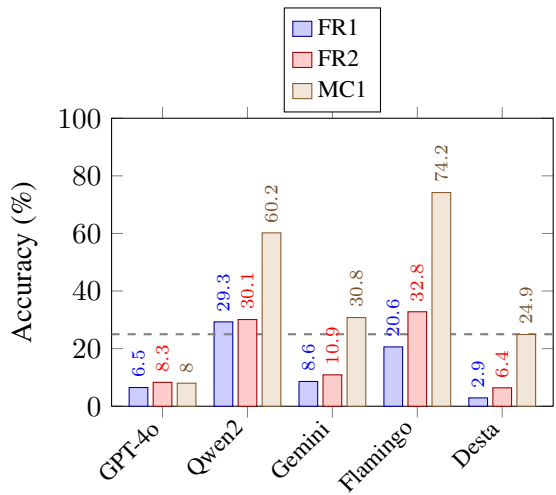
\begin{figure}[ht]
\centering
\begin{tikzpicture}
\begin{axis}[
    height=5.4cm,
    width=0.96\linewidth,
    ybar,
    ymin=0, ymax=100,
    bar width=7pt,
    enlarge x limits=0.15,
    symbolic x coords={GPT-4o, Qwen2, Gemini, Flamingo, Desta},
    xtick=data,
    xticklabel style={rotate=45, anchor=east, font=\small},
    ylabel={Accuracy (\%)},
    legend style={font=\footnotesize, at={(0.5,1.02)}, anchor=south, cells={anchor=west}},
    nodes near coords,
    nodes near coords style={font=\scriptsize, rotate=90, anchor=west},
    legend image code/.code={
        \draw[#1] (0cm,-0.1cm) rectangle (0.25cm,0.1cm);
    },
]

\addplot+[draw=blue!60!black, fill=blue!20] coordinates {
    (GPT-4o,6.5)   +- (0,0.77)
    (Qwen2,29.3)   +- (0,1.41)
    (Gemini,8.6)   +- (0,0.87)
    (Flamingo,20.6)   +- (0,1.25)
    (Desta,2.9)   +- (0,0.52)
};
\addlegendentry{FR1}

\draw[gray, dashed, thick] (rel axis cs:0,0.25) -- (rel axis cs:1,0.25);

\addplot+[draw=red!60!black, fill=red!20] coordinates {
    (GPT-4o,8.3)   +- (0,0.86)
    (Qwen2,30.1)   +- (0,1.42)
    (Gemini,10.9)  +- (0,0.97)
    (Flamingo,32.8)   +- (0,1.45)
    (Desta,6.4)    +- (0,0.76)
};
\addlegendentry{FR2}

\addplot+[draw=brown!70!black, fill=brown!20] coordinates {
    (GPT-4o,8.0)   +- (0,0.84)
    (Qwen2,60.2)   +- (0,1.52)
    (Gemini,30.8)  +- (0,1.43)
    (Flamingo,74.2)   +- (0,1.36)
    (Desta,24.9)   +- (0,1.34)
};
\addlegendentry{MC1}

\end{axis}
\end{tikzpicture}
\caption{\textbf{Ambient\textendash sound accuracy across models} ($N{=}4000$ clips). “Specific Type*” treats FR1\textendash correct as FR2\textendash correct. Baselines for simple chance are 25\% (MC; 1 of 4).}
\label{fig:ambient_accuracy_percent}
\end{figure}

\paragraph{Latency.}
For models with timing logs, we sum FR1 + FR2 (only if FR1 failed) + MC1. Flamingo is fast (median \textbf{2.26}s; p90 2.73s), while Desta is slow (median \textbf{15.61}s). GPT\textendash4o and Gemini are omitted because they are API-based models.

\paragraph{} Models seldom spontaneously mention background noise (FR1) but perform much better with an explicit prompt (FR2) or choices (MC1). Flamingo leads on FR2/MC1; Qwen2 leads FR1. Direct, specific questions markedly improve performance.

\subsection{Noise Localization}
\label{sec:distance-results}
We test whether models detect motion via amplitude envelopes applied to ESC-50 clips (Approaching ↑amp, Receding ↓amp, Oscillating sinusoid). Each clip gets two prompts: FR1 (General) free-text description; FR2 (Direction) explicit location/motion query. A response is correct if it names the ground-truth motion.

\paragraph{FR1: General Description}
\label{sec:motion-FR1}
FR1 remains challenging for all models. Per-class, Receding is consistently the easiest in FR1 (e.g., Gemini 27.4\%), while Oscillating remains near floor for all ($\leq$10\%).

\begin{table}[H]
\centering
\footnotesize
\setlength{\tabcolsep}{6pt}
\begin{tabular}{lccc}
\toprule
\textbf{Model} & \textbf{Correct} & \textbf{Accuracy (\%)} & \textbf{95\% CI} \\
\midrule
Gemini  & 981 & 16.35 & [15.4, 17.3] \\
GPT-4o  & 658  & 10.97 & [10.20, 11.78] \\
Flamingo& 481  &  8.02 & [ 7.49,  8.58] \\
Desta   & 362   &  6.03 & [ 5.51,  6.65] \\
Qwen    & 447 &  7.45 & [ 6.81,  8.14] \\
\bottomrule
\end{tabular}
\caption{\textbf{Motion FR1 (general description)} — accuracy with Wilson 95\% CIs; $N=6000$ per model.}
\label{tab:motion-FR1}
\end{table}

\paragraph{FR2: Direction}
When we allow the “FR1-already correct $\Rightarrow$ auto-correct on FR2” rule and ask direction explicitly, scores jump for the best models:

\begin{table}[H]
\centering
\footnotesize
\setlength{\tabcolsep}{6pt}
\begin{tabular}{lccc}
\toprule
\textbf{Model} & \textbf{Correct} & \textbf{Accuracy (\%)} & \textbf{95\% CI} \\
\midrule
Qwen    & 3434  & 57.23 & [55.98, 58.48] \\
GPT-4o  & 2938  & 48.97 & [47.70, 50.23] \\
Gemini  & 1934 & 32.23 & [31.06, 33.43] \\
Desta   & 1344  & 22.40 & [21.36, 23.47] \\
Flamingo& 1209  & 20.15 & [19.11, 21.23] \\
\bottomrule
\end{tabular}
\caption{\textbf{Motion FR2 (direction)} — accuracy with Wilson 95\% CIs; $N=6000$ per model. FR2 counts FR1-correct as FR2-correct by design.}
\label{tab:motion-FR2}
\end{table}

Asking directly (FR2) helps: $\Delta$(FR2–FR1) = +38.0 pts (GPT-4o), +49.8 pts (Qwen), +15.9 (Gemini), +12.1 (Flamingo), +16.4 (Desta). Models struggle to spontaneously volunteer motion cues in free text (FR1) but can often answer when asked explicitly (FR2). Oscillation remains an unsolved regime.

\subsubsection{Latency}
We report “effective” per-clip latency = FR1 latency + FR2 latency if FR1 failed (0 if FR2 not asked). Logged (local) medians: Flamingo 2.32\,s; Qwen 6.04\,s; Desta 14.53\,s. API models (GPT-4o, Gemini) were not timed in this run. A full latency table is provided in \hyperref[app:motion-stats]{Appendix~\ref*{app:motion-stats}}.

\subsection{Cross-Linguistic Evaluation}
\label{sec:crosslingual-results}
\begin{figure}[ht]
\centering
\includegraphics[width=\linewidth]{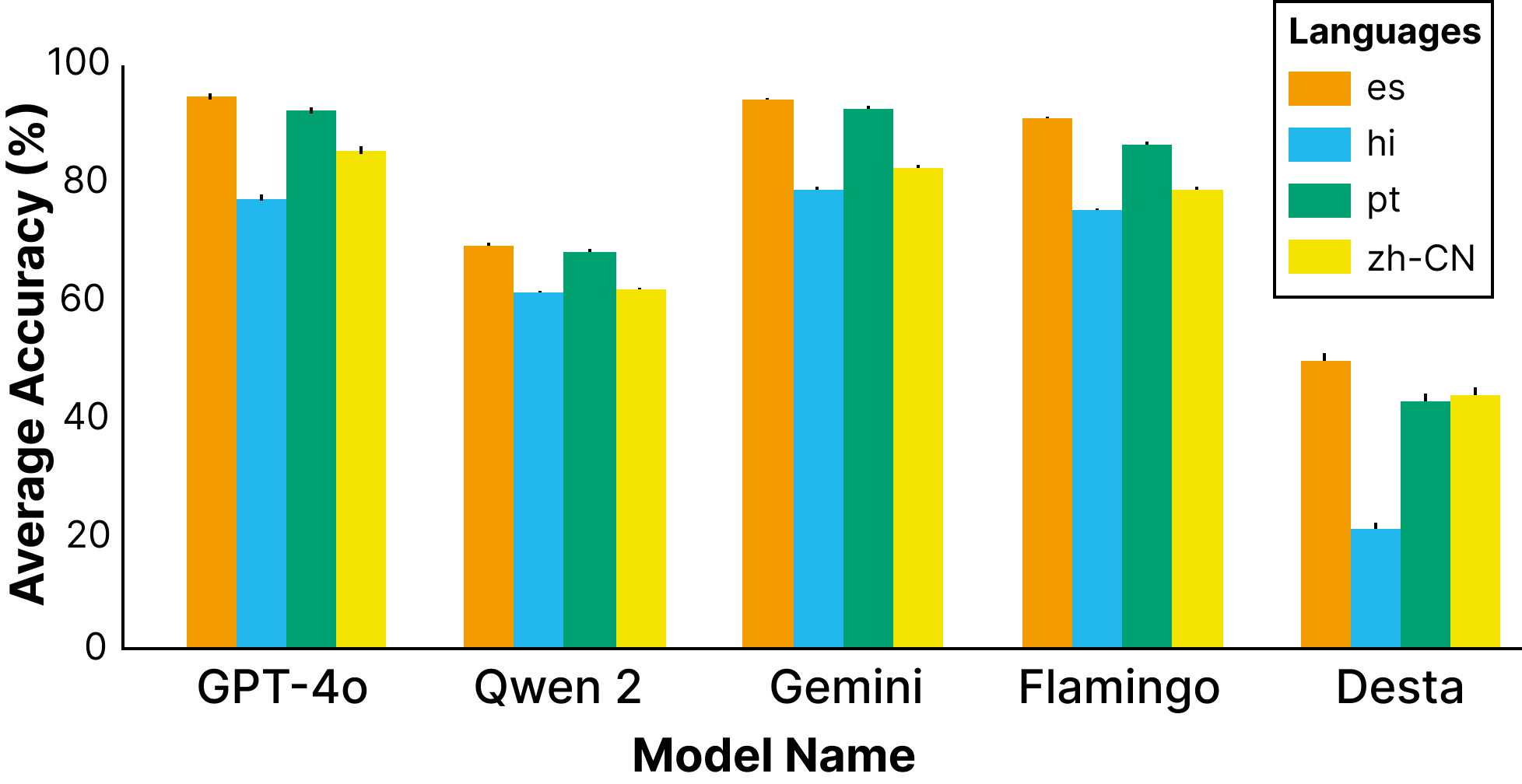}
\caption{\textbf{Cross-linguistic transcription accuracy (mean $\pm$ 95\% CI) across models and languages.} Languages: Spanish (\texttt{es}), Hindi (\texttt{hi}), Portuguese (\texttt{pt}), Mandarin Chinese (\texttt{zh-CN}). Per-language clip counts are $N_\text{es}{=}1010$, $N_\text{hi}{=}1034$, $N_\text{pt}{=}1052$, $N_{\text{zh-CN}}{=}884$. Error bars are normal-approximation CIs over per-clip similarity scores.}
\label{fig:cross_linguistic_accuracy_models}
\end{figure}

We measure transcription \emph{similarity} (0–1; reported as \%) on multilingual DailyTalk clips (construction in \hyperref[sec:crosslingual]{Section~\ref*{sec:crosslingual}}). The 95\% CIs on per-language means are very narrow (typically $<\!1$ percentage point), so they can be hard to discern on the bars in Figure~\ref{fig:cross_linguistic_accuracy_models}. For completeness, the exact CI ranges for every model~$\times$~language are reported in \hyperref[app:xl-stats]{Appendix~\ref*{app:xl-stats}}. 

GPT\textendash4o and Gemini trade the lead by language (Spanish/Portuguese near-ties; Hindi → Gemini; Mandarin → GPT\textendash4o). Flamingo trails the leaders by a small margin on Spanish/Portuguese; Qwen2 is mid-pack; Desta is lowest. None of the models are at ceiling on Mandarin.

Local (open\textendash weights) models only: Flamingo median \textbf{0.92}s (p90 1.23s), Qwen2 \textbf{1.41}s (p90 1.92s), Desta \textbf{4.93}s (p90 10.13s). API models (GPT\textendash4o, Gemini) were not timed in this run.

\subsection{Vocal Characterizers}
\label{sec:vocal-results}
Non-speech vocalizations (cough, cry, laugh, sneeze, yawn) test recognition of acoustic form without relying on linguistic content. We evaluate \textbf{five-way multiple choice} on \textbf{4{,}006} clips (632 cough, 1{,}791 cry, 1{,}133 laugh, 236 sneeze, 214 yawn). We report \textbf{mean\,$\pm$\,95\% CI} to keep tables compact. For a simple chance reference, the five-way baseline is 20\%.

\begin{table}[H]
  \centering
  \footnotesize
  \setlength{\tabcolsep}{5pt}
  \begin{tabular}{lccccc}
    \toprule
    \textbf{Model} & \textbf{Cough} & \textbf{Cry} & \textbf{Laugh} & \textbf{Sneeze} & \textbf{Yawn}\\
    \midrule
    GPT-4o   & 91.5 & 72.1 & 87.6 & 71.6 & 87.9\\
       & $\pm$\,2.2 & $\pm$\,2.1 & $\pm$\,1.9 & $\pm$\,5.8 & $\pm$\,4.4\\
    Gemini   & 48.1 & 65.2 & 65.3 & 14.0 & 7.5\\
     & $\pm$\,3.9 & $\pm$\,2.2 & $\pm$\,2.8 & $\pm$\,4.5 & $\pm$\,3.6\\
    Qwen2    & 74.1 & 69.0 & 97.2 & 79.2 & 86.4\\
      & $\pm$\,3.4 &$\pm$\,2.2 & $\pm$\,1.0 & $\pm$\,5.2 & $\pm$\,4.6\\
    Flamingo & 95.6 & 90.7 & 98.0 & 94.9 & 0.0\\
     & $\pm$\,1.6 & $\pm$\,1.4 & $\pm$\,0.8 & $\pm$\,2.9 & $\pm$\,0.9\\
    Desta    & 26.3 & 11.2 & 64.1 & 3.4 & 7.0\\
     & $\pm$\,3.4 & $\pm$\,1.5 & $\pm$\,2.8 & $\pm$\,2.4 & $\pm$\,3.5\\
    \bottomrule
  \end{tabular}
  \caption{\textbf{Vocal characterizer accuracy} (percent). Entries are mean\,$\pm$\,95\% CI.}
\end{table}

Effective latency sums description\,+\,MC times (local models only): Flamingo \textbf{0.80}s (p90 0.97), Qwen2 \textbf{1.22}s (p90 1.74), Desta \textbf{6.84}s (p90 62.52). API models (GPT\textendash4o, Gemini) were not timed in this run. Flamingo leads on cough/cry/laugh/sneeze but fails on yawn; GPT\textendash4o and Qwen2 are strong across all five; Gemini is mixed and below chance on sneeze/yawn; Desta trails. All models exceed the 20\% chance level overall.

\subsection{Natural Samples}
To assess ecological validity, we additionally evaluate a small naturally occurring human recording dataset\footnote{Released along with primary dataset.} of $N{=}20$ clips per task, matched to the same labels and question formats as the synthetic benchmarks (details on construction in \hyperref[app:human-samples-construct]{Section~\ref*{app:human-samples-construct}}). Rather than synthetically modifying audio, these samples are selected from existing datasets to satisfy the acoustic characteristics required by each task. As a result, the human clips exhibit more natural variability in background noise, speaker position, and language mixing.

\begin{figure*}[!t]
  \centering
  \includegraphics[width=\linewidth]{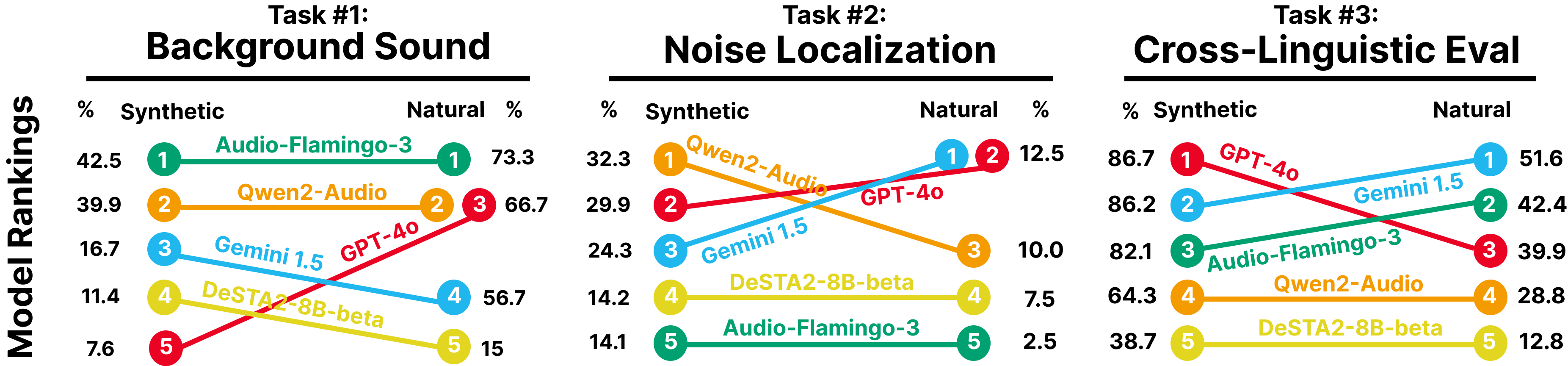}
  \caption{\textbf{Model Rankings on Synthetic and Natural Data.} Although per-task model rankings differ between synthetic and human-recorded evaluations, models that perform well on synthetic data tend to remain strong on average across natural recordings.}
\label{fig:model-rank}
\end{figure*}

Following the analysis of ranking stability in \citet{salaudeen2025imagenotcontrastimagenetpreserves}, we examine whether relative model ordering is preserved across synthetic and human-labeled data. Across tasks, while absolute accuracy ranges differ from synthetic results, relative performance patterns are generally preserved, as seen in \hyperref[fig:model-rank]{Figure~\ref*{fig:model-rank}}: models that perform well on synthetic data tend to remain strong on human recordings, and weaker models continue to underperform.

Across tasks, absolute performance on human-recorded clips differs from synthetic benchmarks, yet overall relative model rankings is largely preserved (Figure~\ref{fig:model-rank}), suggesting that synthetic evaluations capture meaningful comparative trends. Models perform better on background sound understanding in natural audio, likely because salient events occur intermittently rather than being fully mixed, while noise localization degrades due to subtler spatial cues and the presence of speech that draws models toward transcription over spatial reasoning. Cross-linguistic accuracy similarly declines in human recordings, reflecting greater linguistic variability and limited code-switching, which causes models to remain anchored to the dominant language despite embedded English content. We do not include a separate human-validation split for vocal characterizers, as these clips in our dataset are directly from real recordings rather than synthetically generated.

\paragraph{}Full details and numbers for each of the human validation tasks are in \hyperref[app:human-lab-t1]{Section~\ref*{app:human-lab-t1}}.

\subsection{Error Analysis}
\label{sec:error-analysis}

\begin{table}[H]
\centering
\footnotesize
\setlength{\tabcolsep}{5pt}
\renewcommand{\arraystretch}{1.1}
\begin{tabular}{@{}lcccc@{}}
\toprule
\textbf{Error type} & \textbf{\shortstack{Bkgd.\\ Sound}} & \textbf{\shortstack{Noise \\Loc.}} & \textbf{\shortstack{Cross \\Ling.}} & \textbf{\shortstack{Vocal \\ chars.} }\\
\midrule
Omission                & 18\% & 43\% & -- & 1\% \\
Over-general label      &  4\% &  --  & -- & -- \\
Misattribution          &  3\% & 21\% & -- & 94\% \\
Direction swap          &  --  &  8\% & -- & -- \\
Dropping language       & -- & -- & 64\% & -- \\
Partial re-translation  & -- & -- & 36\% & -- \\
Noise override          & 75\% & -- & -- & -- \\
\bottomrule
\end{tabular}
\caption{\textbf{Distribution of error types by task} (percent of error cases in the aggregated dataset).  Entries are percentages conditioned on an error in that task, not on all clips.  Cells marked “--” denote that the error category is not applicable to that task.}
\label{tab:error-distribution}
\end{table}

Beyond aggregate accuracy, we perform a structured error analysis to understand how models fail on each task. For each task, we drew a stratified random sample of error cases across models (10 clips per model per task; $N{=}200$ total). We then developed an error taxonomy on this set and labeled the full set of samples using this schema. Our labels include (i) \emph{omission} (the target event is never mentioned), (ii) \emph{over-general labels} (e.g., ``traffic'' instead of a specific ESC-50 class), (iii) \emph{misattribution} (wrong specific class), (iv) \emph{direction swap} (approach$\leftrightarrow$recede), (v) \emph{dropping language} (dropping non-English spans), (vi) \emph{partial re-translation} (translating only the foreign span), and (vii) \emph{noise override} (focusing on speech content of the audio snippet and ignoring the background cues). Error types are not mutually exclusive.

The dominant failure for \texttt{background} \texttt{sound} is \emph{noise override}: in 75\% of wrong free responses, models transcribe the speech but never mention the background event. Plain omissions account for 18\% of errors, over-general labels (e.g., ``background noise,'' ``traffic'') for 4\%, and misattributions for 3\%.

For \texttt{noise} \texttt{localization}, the main issues are omission (43\%; no motion described), \emph{oscillation collapse} (25\%; periodic loud--soft patterns reduced to ``volume changes''), and misattributed motion type (21\%). Direction swaps (approach$\leftrightarrow$recede) contribute 8\% and short-horizon listening 2\%.

In \texttt{cross-lingual} transcription, 64\% of errors are monolingual normalization, where non-English spans are dropped and replaced with fluent English, and 36\% are partial re-translation, where only the foreign fragment is translated despite instructions to preserve code-mixing.

For \texttt{vocal} \texttt{characterizers}, 94\% of errors are misattributions between non-speech categories (e.g., yawn vs.\ sigh), with the mention of vocal characterizers outside of our five labels at 5\% and omissions at about 1\%.

Taken together, these patterns show a bias toward foreground speech over background events, limited temporal reasoning for motion, normalization of multilingual spans, and unstable labeling of non-speech vocal cues.

\section{Discussion}
\label{sec:discussion}

SCENEBench probes LALMs' audio capabilities by assessing whether they can detect salient background events, reason about motion, preserve multilingual spans, and recognize non-speech vocalizations in realistic settings. Our results are mixed.

Across tasks, models approach or surpass chance on most multiple-choice formats but routinely miss the phenomena that matter in assistive and industrial use cases. In \texttt{background sound understanding}, omission dominates: models almost never volunteer background events in free text and only partially recover when pushed with targeted questions or options. In \texttt{noise localization}, models show substantial gains when we explicitly ask about motion, but still collapse oscillatory envelopes and over-weight the end of the clip. In \texttt{cross-linguistic transcription}, the most common error is to normalize away the non-English spans entirely. In \texttt{vocal characterizers}, lexical content overrides paralinguistic cues even when the non-lexical event (e.g., a yawn) is the label of interest.

These patterns suggest that current LALMs are optimized for \emph{what is said} (ASR and captioning) rather than \emph{how it is said} or \emph{what else is happening} in the scene. This is not surprising: most public audio benchmarks focus on clean speech, single-source environmental clips, or music classification, and are designed around tasks that are easy to grade automatically. As a result, it is possible for a model to score well on existing suites while still failing on the basic building blocks of audio understanding.

So why have these gaps been overlooked? Our analysis suggests several structural reasons why these capabilities have received less attention. First, collecting and annotating overlapping events (speech with background, motion, and paralinguistic cues) is much harder than curating clean, single-label clips; therefore, most widely used corpora simply do not contain them. Additionally, common metrics such as word-error rate, caption BLEU, or single-label accuracy reward lexical fidelity and generic scene tags, but do not penalize models for dropping sirens, collapsing motion, or normalizing away multilingual spans. SCENEBench is designed around exactly these ``inconvenient'' cases, not to replace existing suites, but to make it harder to claim comprehensive audio understanding without addressing them.

One way to improve performance on these tasks is to focus on targeted data and training objectives. For background sound understanding one method of improving existing models is exposing models to speech–noise mixtures with descriptions of the background sound. Specifically, we could include information regarding (i) the presence of a background class and (ii) name it, providing the model with hard negatives where the foreground narrative is correct but the background label is wrong, since LALMs perform well at ASR \citep{chu2024qwen2audio,AudioBench2024,AIRBench2024}. For localization, train on clips with known object movement; add objectives that classify approach/recede/oscillate, and ensure that the model integrates evidence over time rather than over-weighting the final seconds. For multilingual spans, instruction- and preference-tuning on span-annotated transcripts, using contrastive pairs where the only difference is whether to keep or translate the foreign segment. For paralinguistics, fine-tune or introduce data that includes short non-lexical events embedded in speech. Across tasks, use multi-task fine-tuning that combines these objectives with standard ASR/captioning, ensuring lexical fidelity while the model also learns to attend to the features of audio.

To track whether these targeted changes actually help in practice, our benchmark provides the corresponding evaluation signals. In summary, the suite identifies where current models excel and where they falter, providing a concise, reproducible testbed to guide training and model design toward parsing audio, not just speech.

\subsection{Future Work}

To enhance coverage, there are three improvements that can be prioritized in future work:

\begin{compactitem}
  \item \textbf{Natural code-switching data.} Replace the synthetically generated data in our third task with a small, human-recorded corpus across multiple language pairs to validate (or revise) conclusions from the synthetic set and better reflect real switching behavior.
  \item \textbf{Realistic acoustics.} Move beyond equal-loudness overlays and synthetic motion by (i) sweeping speech–background SNRs (e.g., $-10$ to $+10$\,dB) and (ii) adding field recordings with moving sources (sirens/vehicles/machinery) to test approach/recede under Doppler, occlusion, and moving-listener cases.
  \item \textbf{Stronger baselines.} Include non–LALM pipelines to contextualize results (e.g., speech separation $\rightarrow$ ESC\textendash50 classifier for \texttt{background sound understanding}); this sets a clear “what classical methods achieve” line for robustness comparisons and allows us to indicate whether results indicate task difficulty or model limitations.
\end{compactitem}

\section {Conclusion}

Overall, \textbf{SCENEBench} surfaces concrete failure modes—omission, over-general labels, misattribution, short-horizon listening, and language-prior dominance—that are obscured in cleaner, single-event benchmarks. We provide task-appropriate chance baselines, report tight confidence intervals enabled by large $N$, and keep full statistics in \hyperref[app:stats]{Appendix~\ref*{app:stats}}. Our hope is that these evaluations help steer model development toward the capabilities demanded in accessibility and industrial safety.

\section{Limitations}
\label{sec:limitations}
Our study has several limitations. First, the mixture design relies on controlled levels of speech–noise overlays. While equal-level mixing provides experimental control, it may not reflect signal-to-noise ratio distributions encountered in the wild. Second, some of the upstream corpora contain weak or imperfect annotations. Residual label errors can influence ceiling estimates and increase per-class confusions. Third, closed-model APIs restrict fine-grained latency profiling and ablation studies. As a result, we report timer latency measures wherever possible. Finally, the multilingual speech setup, which uses a TTS pipeline with back-translation filtering, improves consistency but does not fully capture the spontaneity of real human speech.

\section{Ethical Considerations}
We also highlight several ethical considerations. To reduce risks associated with reductive “emotion AI,” we avoid direct emotion inference and instead focus on paralinguistic events. All speech data, including synthetic mixtures, must respect licensing constraints, and any human recordings require explicit consent and data minimization. Accessibility risks must be considered carefully: in safety-critical contexts like siren detection, both missed detections and false alarms can carry distinct harms. Systems should therefore expose calibrated uncertainty and provide fallback behaviors. Finally, fairness is essential. Benchmarks should broaden their coverage of accents, languages, and recording conditions to reduce disparate error rates across user groups.

Disclosure: LLMs were used to refine the text and the tables. 
\\\\
SK is partially supported by NSF 2046795 and 2205329, IES R305C240046, ARPA-H, the MacArthur Foundation, Schmidt Sciences, Stanford HAI, RAISE Health, OpenAI, Microsoft, and Google.

\bibliography{custom}

@misc{salaudeen2025imagenotcontrastimagenetpreserves,
      title={ImageNot: A contrast with ImageNet preserves model rankings}, 
      author={Olawale Salaudeen and Moritz Hardt},
      year={2025},
      eprint={2404.02112},
      archivePrefix={arXiv},
      primaryClass={cs.LG},
      url={https://arxiv.org/abs/2404.02112}, 
}

@inproceedings{Zhang_2025,
   title={LMMs-Eval: Reality Check on the Evaluation of Large Multimodal Models},
   url={http://dx.doi.org/10.18653/v1/2025.findings-naacl.51},
   DOI={10.18653/v1/2025.findings-naacl.51},
   booktitle={Findings of the Association for Computational Linguistics: NAACL 2025},
   publisher={Association for Computational Linguistics},
   author={Zhang, Kaichen and Li, Bo and Zhang, Peiyuan and Pu, Fanyi and Cahyono, Joshua Adrian and Hu, Kairui and Liu, Shuai and Zhang, Yuanhan and Yang, Jingkang and Li, Chunyuan and Liu, Ziwei},
   year={2025},
   pages={881–916} }

@misc{surapaneni2025auharnessopensourcetoolkitholistic,
      title={AU-Harness: An Open-Source Toolkit for Holistic Evaluation of Audio LLMs}, 
      author={Sidharth Surapaneni and Hoang Nguyen and Jash Mehta and Aman Tiwari and Oluwanifemi Bamgbose and Akshay Kalkunte and Sai Rajeswar and Sathwik Tejaswi Madhusudhan},
      year={2025},
      eprint={2509.08031},
      archivePrefix={arXiv},
      primaryClass={cs.SD},
      url={https://arxiv.org/abs/2509.08031}, 
}

@inproceedings{panayotov2015librispeech,
  title={Librispeech: an ASR corpus based on public domain audio books},
  author={Panayotov, Vassil and Chen, Guoguo and Povey, Daniel and Khudanpur, Sanjeev},
  booktitle={Acoustics, Speech and Signal Processing (ICASSP), 2015 IEEE International Conference on},
  pages={5206--5210},
  year={2015},
  organization={IEEE}
}

@misc{soketlabs_coshe_eval_2025,
  title        = {CoSHE-Eval: A Code-Switching ASR Benchmark for Hindi-English Speech},
  author       = {Soket Labs},
  year         = {2025},
  howpublished = {Hugging Face Dataset},
  url          = {https://huggingface.co/datasets/soketlabs/CoSHE-Eval}
}

@inproceedings{seame2015,
  title     = {SEAME: a Mandarin-English code-switching speech corpus in south-east asia},
  author    = {Dau-Cheng Lyu and Tien-Ping Tan and Eng Siong Chng and Haizhou Li},
  year      = {2010},
  booktitle = {Interspeech 2010},
  pages     = {1986--1989},
  doi       = {10.21437/Interspeech.2010-563},
  issn      = {2958-1796},
}

@inproceedings{Piczak2015ESC50,
  author    = {Karol J. Piczak},
  title     = {{ESC-50}: Dataset for Environmental Sound Classification},
  booktitle = {Proc.\ ACM MM},
  year      = {2015},
  pages     = {1015--1018},
  doi       = {10.1145/2733373.2806390}
}

@misc{openai2024gpt4o,
  author       = {{OpenAI}},
  title        = {{GPT‑4o System Card}},
  howpublished = {\url{https://cdn.openai.com/gpt-4o-system-card.pdf}},
  month        = {August},
  year         = {2024},
  note         = {Accessed 24~July~2025}
}

@misc{chu2024qwen2audio,
      title={Qwen2-Audio Technical Report}, 
      author={Yunfei Chu and Jin Xu and Qian Yang and Haojie Wei and Xipin Wei and Zhifang Guo and Yichong Leng and Yuanjun Lv and Jinzheng He and Junyang Lin and Chang Zhou and Jingren Zhou},
      year={2024},
      eprint={2407.10759},
      archivePrefix={arXiv},
      primaryClass={eess.AS},
      url={https://arxiv.org/abs/2407.10759}, 
}

@article{Chin2023,
  author    = {Chiun-Li Chin and Chia-Chun Lin and Jing-Wen Wang and Wei-Cheng Chin and Yu-Hsiang Chen and Sheng-Wen Chen and Po-Cheng Hsieh},
  title     = {A Wearable Assistant Device for the Hearing Impaired to Recognize Emergency Vehicle Sirens with Edge Computing},
  journal   = {Sensors},
  year      = {2023},
  volume    = {23},
  number    = {17},
  pages     = {7454},
  doi       = {10.3390/s23177454},
  url       = {https://doi.org/10.3390/s23177454}
}

@inproceedings{Salem2023,
  author    = {Osman Salem and Ahmed Mehaoua and Raouf Boutaba},
  title     = {The Sight for Hearing: An IoT-Based System to Assist Drivers with Hearing Disability},
  booktitle = {2023 IEEE Symposium on Computers and Communications (ISCC)},
  year      = {2023},
  doi       = {10.1109/ISCC58397.2023.10218250},
  publisher = {IEEE},
  note      = {Available at\url{https://rboutaba.cs.uwaterloo.ca/Papers/Conferences/2023/SalemIoT2023.pdf}}
}

@misc{AIMultiple2025,
  author       = {Cem Dilmegani},
  title        = {Top 4 Speech Recognition Challenges \& Solutions in 2025},
  year         = {2025},
  howpublished = {\url{https://research.aimultiple.com/speech-recognition-challenges/}},
  note         = {Accessed: 2025-07-24}
}

@article{AudioBench2024,
  author    = {Bin Wang and Xunlong Zou and Geyu Lin and Shuo Sun and Zhuohan Liu and Wenyu Zhang and Zhengyuan Liu and AiTi Aw and Nancy F. Chen},
  title     = {AudioBench: A Universal Benchmark for Audio Large Language Models},
  journal   = {arXiv preprint arXiv:2406.16020},
  year      = {2025},
  note      = {Last revised May~2025},
  url       = {https://arxiv.org/abs/2406.16020}
}

@article{MMAU2024,
  author    = {S. Sakshi and Utkarsh Tyagi and Sonal Kumar and Ashish Seth and Ramaneswaran Selvakumar and Oriol Nieto and Ramani Duraiswami and Sreyan Ghosh and Dinesh Manocha},
  title     = {MMAU: A Massive Multi-Task Audio Understanding and Reasoning Benchmark},
  journal   = {arXiv preprint arXiv:2410.19168},
  year      = {2024},
  url       = {https://arxiv.org/abs/2410.19168}
}

@article{AIRBench2024,
  author    = {Qian Yang and Jin Xu and Wenrui Liu and Yunfei Chu and Ziyue Jiang and Xiaohuan Zhou and Yichong Leng and Yuanjun Lv and Zhou Zhao and Chang Zhou and Jingren Zhou},
  title     = {AIR-Bench: Benchmarking Large Audio-Language Models via Generative Comprehension},
  journal   = {arXiv preprint arXiv:2402.07729},
  year      = {2024},
  note      = {Accepted to ACL 2024},
  url       = {https://arxiv.org/abs/2402.07729}
}

@article{MARBLE2023,
  author    = {Ruibin Yuan and Yinghao Ma and Yizhi Li and Ge Zhang and Xingran Chen and Hanzhi Yin and Le Zhuo and Yiqi Liu and Jiawen Huang and Zeyue Tian and Binyue Deng and Ningzhi Wang and Chenghua Lin and Emmanouil Benetos and Anton Ragni and Norbert Gyenge and Roger Dannenberg and Wenhu Chen and Gus Xia and Wei Xue and Si Liu and Shi Wang and Ruibo Liu and Yike Guo and Jie Fu},
  title     = {MARBLE: Music Audio Representation Benchmark for Universal Evaluation},
  journal   = {arXiv preprint arXiv:2306.10548},
  year      = {2023},
  note      = {Camera-ready version for NeurIPS 2023},
  url       = {https://arxiv.org/abs/2306.10548}
}

@article{ClothoAQA2022,
  author    = {Samuel Lipping and Parthasaarathy Sudarsanam and Konstantinos Drossos and Tuomas Virtanen},
  title     = {Clotho-AQA: A Crowdsourced Dataset for Audio Question Answering},
  journal   = {arXiv preprint arXiv:2204.09634},
  year      = {2022},
  url       = {https://arxiv.org/abs/2204.09634}
}

@article{SoundCheck2023,
  author    = {William Agnew and Julia Barnett and Annie Chu and Rachel Hong and Michael Feffer and Robin Netzorg and Harry H. Jiang and Ezra Awumey and Sauvik Das},
  title     = {Sound Check: Auditing Audio Datasets},
  journal   = {arXiv preprint arXiv:2410.13114},
  year      = {2024},
  url       = {https://arxiv.org/abs/2410.13114}
}

@article{SONAR2023,
  author    = {Xiang Li and Pin-Yu Chen and Wenqi Wei},
  title     = {SONAR: A Synthetic AI-Audio Detection Framework and Benchmark},
  year={2025},
  eprint={2410.04324},
  archivePrefix={arXiv},
  primaryClass={cs.SD},
  url={https://arxiv.org/abs/2410.04324}
}

@misc{CAVA2025,
  title = {CAVA: Comprehensive Assessment of Voice Assistants},
  author = {Held, Will and Ryan, Michael J. and Shrivastava, Aditya and Khan, Ali Sartaz and Ziems, Caleb and Li, Ella and Bartelds, Martijn and Sun, Michael and Li, Tan and Gan, Woody and Yang, Diyi},
  year = {2025},
  url = {https://talkarena.org/cava},
  howpublished = {\url{https://github.com/SALT-NLP/CAVA}},
  note = {A benchmark for evaluating large audio models (LAMs) capabilities across six domains: turn taking, instruction following, function calling, tone awareness, safety, and latency}
}

@misc{goel2025audioflamingo3advancing,
      title={Audio Flamingo 3: Advancing Audio Intelligence with Fully Open Large Audio Language Models}, 
      author={Arushi Goel and Sreyan Ghosh and Jaehyeon Kim and Sonal Kumar and Zhifeng Kong and Sang-gil Lee and Chao-Han Huck Yang and Ramani Duraiswami and Dinesh Manocha and Rafael Valle and Bryan Catanzaro},
      year={2025},
      eprint={2507.08128},
      archivePrefix={arXiv},
      primaryClass={cs.SD},
      url={https://arxiv.org/abs/2507.08128}, 
}

@article{lu2024developing,
  title={Developing Instruction-Following Speech Language Model Without Speech Instruction-Tuning Data},
  author={Lu, Ke-Han and Chen, Zhehuai and Fu, Szu-Wei and Yang, Chao-Han Huck and Balam, Jagadeesh and Ginsburg, Boris and Wang, Yu-Chiang Frank and Lee, Hung-yi},
  journal={arXiv preprint arXiv:2409.20007},
  year={2024}
}

@inproceedings{koizumi2019toyadmos,
  author    = {Yuma Koizumi and Shoichiro Saito and Hisashi Uematsu and Noboru Harada and Keisuke Imoto},
  title     = {ToyADMOS: A dataset of miniature-machine operating sounds for anomalous sound detection},
  booktitle = {Proc. IEEE WASPAA},
  year      = {2019}
}

@inproceedings{purohit2019mimii,
  author    = {Harsh Purohit and Ryo Tanabe and Kohei Ichige and Tomoya Nishida and Takashi Endo and Yuma Koizumi and Noboru Harada and Masahiro Yasuda and Satoru Tamura},
  title     = {MIMII Dataset: Sound dataset for malfunctioning industrial machine investigation and inspection},
  booktitle = {Proc. DCASE Workshop},
  year      = {2019},
  note      = {arXiv:1909.09347}
}

@misc{dohi2022mimiidg,
      title={MIMII DG: Sound Dataset for Malfunctioning Industrial Machine Investigation and Inspection for Domain Generalization Task}, 
      author={Kota Dohi and Tomoya Nishida and Harsh Purohit and Ryo Tanabe and Takashi Endo and Masaaki Yamamoto and Yuki Nikaido and Yohei Kawaguchi},
      year={2022},
      eprint={2205.13879},
      archivePrefix={arXiv},
      primaryClass={cs.SD},
      url={https://arxiv.org/abs/2205.13879}, 
}

@article{orlandic2021coughvid,
  author  = {Orlandic, Lara and Teijeiro, Tomislav and Atienza, David},
  title   = {The COUGHVID Crowdsourcing Dataset: A corpus for COVID-19 cough classification},
  journal = {Scientific Data},
  year    = {2021},
  volume  = {8},
  number  = {1},
  pages   = {156}
}

@article{sharma2020coswara,
  author  = {Bhattacharya, Debarpan and
             Sharma, Neeraj Kumar and
             Dutta, Debottam and
             Chetupalli, Srikanth Raj and
             Mote, Pravin and
             Ganapathy, Sriram and
             Chandrakiran, C. and
             Nori, Sahiti and
             Suhail, K. K. and
             Gonuguntla, Sadhana and
             Alagesan, Murali},
  title   = {Coswara: A respiratory sounds and symptoms dataset for remote screening of {SARS-CoV-2} infection},
  journal = {Scientific Data},
  year    = {2023},
  volume  = {10},
  number  = {1},
  pages   = {397},
  doi     = {10.1038/s41597-023-02266-0},
  url     = {https://doi.org/10.1038/s41597-023-02266-0},
}

@inproceedings{schuller2011compareSleepiness,
  author    = {Schuller, Björn and Steidl, Stefan and Batliner, Anton and Schiel, Florian and Krajewski, Jarek},
  title     = {The INTERSPEECH 2011 Computational Paralinguistics Challenge: Intoxication, Sleepiness, and Age},
  booktitle = {Proc. INTERSPEECH},
  year      = {2011}
}

@misc{mesaros2018dcaseDomestic,
      title={DCASE 2018 Challenge - Task 5: Monitoring of domestic activities based on multi-channel acoustics}, 
      author={Gert Dekkers and Lode Vuegen and Toon van Waterschoot and Bart Vanrumste and Peter Karsmakers},
      year={2018},
      eprint={1807.11246},
      archivePrefix={arXiv},
      primaryClass={eess.AS},
      url={https://arxiv.org/abs/1807.11246}, 
}

@inproceedings{mesaros2017dcaseRare,
  author    = {Mesaros, Annamaria and Heittola, Toni and Benetos, Emmanouil},
  title     = {DCASE 2017 Challenge Task 2: Detection of rare sound events in real-life audio},
  booktitle = {Proc. DCASE Workshop},
  year      = {2017}
}

@misc{SCC2025,
  title={Saudilang Code-Switch Corpus (SCC)},
  author={SDAIA},
  year={2022},
  howpublished={\url{https://www.kaggle.com/datasets/sdaiancai/saudilang-code-switch-corpus-scc}},
  note={CC BY-NC-SA 4.0}
}

@INPROCEEDINGS{gemmeke2017audioset,
  author={Gemmeke, Jort F. and Ellis, Daniel P. W. and Freedman, Dylan and Jansen, Aren and Lawrence, Wade and Moore, R. Channing and Plakal, Manoj and Ritter, Marvin},
  booktitle={2017 IEEE International Conference on Acoustics, Speech and Signal Processing (ICASSP)}, 
  title={Audio Set: An ontology and human-labeled dataset for audio events}, 
  year={2017},
  volume={},
  number={},
  pages={776-780},
  doi={10.1109/ICASSP.2017.7952261}}

@misc{gordon2020soundspaces,
      title={SoundSpaces: Audio-Visual Navigation in 3D Environments}, 
      author={Changan Chen and Unnat Jain and Carl Schissler and Sebastia Vicenc Amengual Gari and Ziad Al-Halah and Vamsi Krishna Ithapu and Philip Robinson and Kristen Grauman},
      year={2020},
      eprint={1912.11474},
      archivePrefix={arXiv},
      primaryClass={cs.CV},
      url={https://arxiv.org/abs/1912.11474}, 
}

@misc{chen2020audiogoal,
      title={Semantic Audio-Visual Navigation}, 
      author={Changan Chen and Ziad Al-Halah and Kristen Grauman},
      year={2021},
      eprint={2012.11583},
      archivePrefix={arXiv},
      primaryClass={cs.CV},
      url={https://arxiv.org/abs/2012.11583}, 
}

@inproceedings{hershey2016deepclustering,
  author    = {Hershey, John R. and Chen, Zhuo and Le Roux, Jonathan and Watanabe, Shinji},
  title     = {Deep Clustering: Discriminative Embeddings for Segmentation and Separation},
  booktitle = {Advances in Neural Information Processing Systems (NeurIPS)},
  year      = {2016},
  note      = {Introduces the WSJ0-2mix evaluation setup for single-channel speech separation}
}

@inproceedings{isik2016deepclustering,
  author    = {Isik, Yusuf and Roux, Jonathan Le and Chen, Zhuo and Watanabe, Shinji and Hershey, John R.},
  title     = {Single-Channel Multi-Speaker Separation Using Deep Clustering},
  booktitle = {Proc.\ Interspeech},
  year      = {2016}
}

@article{cosentino2020librimix,
  author  = {Cosentino, Joris and Pariente, Manuel and Cornell, Samuele and Deleforge, Antoine and Vincent, Emmanuel},
  title   = {LibriMix: An Open-Source Dataset for Generalizable Speech Separation},
  journal = {arXiv preprint arXiv:2005.11262},
  year    = {2020},
  url     = {https://arxiv.org/abs/2005.11262},
  note    = {Includes Libri2Mix and Libri3Mix variants}
}

@inproceedings{lee2023dailytalk,
  author    = {Lee, Keon and Park, Kyumin and Kim, Daeyoung},
  title     = {DailyTalk: Spoken Dialogue Dataset for Conversational Text-to-Speech},
  booktitle = {Proc.\ IEEE ICASSP},
  year      = {2023}
}

@article{stark2021ethics,
  author  = {Stark, Luke and Hoey, Jevin D.},
  title   = {The Ethics of Emotion in AI Systems},
  journal = {Proceedings of the ACM on Human-Computer Interaction},
  year    = {2021},
  volume  = {5},
  number  = {CSCW2},
  pages   = {1--34},
  note    = {Discusses risks of reductive emotion recognition and attendant harms}
}

@misc{nonspeech7k,
  author = {{W4ng1204}},
  title  = {Nonspeech7k: Non-Speech Vocalization Dataset (Cough, Cry, Laugh, Sneeze, Yawn)},
  howpublished = {\url{https://huggingface.co/datasets/W4ng1204/Nonspeech7k}},
  year = {2023},
  note = {Hugging Face dataset; Accessed: 2025-10-06}
}

@misc{capspeechagentdb,
  author       = {{OpenSound}},
  title        = {CapSpeech-AgentDB-Audio: Mumble and Whisper Audio Clips},
  howpublished = {\url{https://huggingface.co/datasets/OpenSound/CapSpeech-AgentDB-Audio}},
  year         = {2024},
  note         = {Hugging Face dataset; Accessed: 2025-10-06}
}

@misc{vocalbursts100,
  author       = {Krishnakalyan, Kalyan},
  title        = {Vocal Bursts Taxonomy 100 Clean (WDS)},
  howpublished = {\url{https://huggingface.co/datasets/krishnakalyan3/vocal_bursts_taxonomy_100_clean_wds}},
  year         = {2023},
  note         = {Accessed: 2025-09-06}
}

@misc{asmrrepo,
  author       = {{Nyuuzyou}},
  title        = {ASMR Whisper Audio Collection},
  howpublished = {\url{https://huggingface.co/datasets/nyuuzyou/asmr}},
  year         = {2022},
  note         = {Hugging Face dataset; Accessed: 2025-10-06}
}

@inproceedings{Jain2019CHI_SoundAwareness,
  author    = {Dhruv Jain and Leah Findlater and Jon E. Froehlich},
  title     = {Exploring Sound Awareness in the Home for People who are Deaf or Hard of Hearing},
  booktitle = {Proceedings of the 2019 CHI Conference on Human Factors in Computing Systems},
  year      = {2019},
  pages     = {94:1--94:13},
  doi       = {10.1145/3290605.3300324}
}

@inproceedings{Bragg2016ASSETS_PersonalizableDetector,
  author    = {Danielle Bragg and Naomi Huynh and Richard E. Ladner},
  title     = {A Personalizable Mobile Sound Detector App Design for Deaf and Hard-of-Hearing Users},
  booktitle = {Proceedings of the 18th International ACM SIGACCESS Conference on Computers and Accessibility (ASSETS)},
  year      = {2016},
  pages     = {3--13},
  doi       = {10.1145/2982142.2982171}
}

@inproceedings{Findlater2019Sound,
  author    = {Leah Findlater and Bonnie Chinh and Dhruv Jain and Jon E. Froehlich and Raja Kushalnagar and Angela Carey Lin},
  title     = {Deaf and Hard-of-Hearing Individuals' Preferences for Wearable and Mobile Sound Awareness Technologies},
  booktitle = {Proceedings of the 2019 CHI Conference on Human Factors in Computing Systems (CHI '19)},
  year      = {2019},
  pages     = {1--13},
  publisher = {ACM},
  address   = {New York, NY, USA},
  doi       = {10.1145/3290605.3300276},
  url       = {https://doi.org/10.1145/3290605.3300276}
}

@inproceedings{VisibleNuances2023,
  author    = {JooYeong Kim and Sooyeon Ahn and Jin{-}Hyuk Hong},
  title     = {Visible Nuances: A Caption System to Visualize Paralinguistic Speech Cues for Deaf and Hard-of-Hearing Individuals},
  booktitle = {Proceedings of the 2023 CHI Conference on Human Factors in Computing Systems (CHI '23)},
  year      = {2023},
  publisher = {ACM},
  address   = {New York, NY, USA},
  doi       = {10.1145/3544548.3581152},
  url       = {https://doi.org/10.1145/3544548.3581152}
}

@inproceedings{SoundNarratives2024,
  author    = {Liang{-}Yuan Wu and Dhruv Jain},
  title     = {SoundNarratives: Rich Auditory Scene Descriptions to Support Deaf and Hard of Hearing People},
  booktitle = {Proceedings of the 27th International ACM SIGACCESS Conference on Computers and Accessibility (ASSETS '25)},
  year      = {2025},
  publisher = {ACM},
  address   = {New York, NY, USA},
  doi       = {10.1145/3663547.3746341},
  note      = {Appears at ASSETS~2025; preprint available},
  url       = {https://doi.org/10.1145/3663547.3746341}
}

@misc{AlibabaQwen2AudioBlog2024,
  author       = {{Alibaba Cloud}},
  title        = {Alibaba Cloud Launches Qwen2-Audio Model to Analyze Speech and Audio},
  year         = {2024},
  url = {https://www.alibabacloud.com/blog/alibaba-cloud-launches-qwen2-audio-model-to-analyze-speech-and-audio_601584},
  note         = {Accessed: 2025-10-06}
}

\appendix
\section{Background Sound Task Prompts}
\label{app:ambient-prompts}

\paragraph{FR1 (Free-response description).}
\begin{quote}
\small
\texttt{Please describe what you hear in this audio file. Be as detailed as possible about all the sounds you can identify.}
\end{quote}

\paragraph{FR2 (Targeted follow-up; asked only if FR1 omitted the background/ambient sound).}
\begin{quote}
\small
\texttt{In your previous response, you didn't mention the background noise. Please specifically describe what background noise or ambient sound you hear in this audio.}
\end{quote}

\paragraph{MC (4-way multiple choice; administered for all clips).}
\begin{quote}
\small
\texttt{You will be given an audio clip. What is the background sound? Choose the correct answer from the following options and reply with ONLY the number (e.g., 1, 2, 3, or 4).}
\end{quote}

\section{Statistical Methods for Background Sound Task}
\label{app:stats}
We compute model-wise 95\% CIs for proportions using the Wilson interval. 
FR1 (Any noise) — GPT\textendash4o: 6.5\% [5.73, 7.27], Qwen2: 29.3\% [27.89, 30.71], Gemini: 8.6\% [7.73, 9.47], Flamingo: 20.6\% [19.35, 21.85], Desta: 2.9\% [2.38, 3.42]. 
FR2 (Specific type*) — GPT\textendash4o: 8.3\% [7.44, 9.16], Qwen2: 30.1\% [28.68, 31.52], Gemini: 10.9\% [9.93, 11.87], Flamingo: 32.8\% [31.35, 34.25], Desta: 6.4\% [5.64, 7.16]. 
MC — GPT\textendash4o: 8.0\% [7.16, 8.84], Qwen2: 60.2\% [58.68, 61.72], Gemini: 30.8\% [29.37, 32.23], Flamingo: 74.2\% [72.84, 75.56], Desta: 24.9\% [23.56, 26.24].

For each model and the multiple choice question (MC1), we test \(\mathrm{H}_0{:}~p = p_{\text{chance}}\) with a one-sample proportion \(z\)-test: \(0.25\) (MC1; 1/4). To control multiplicity across 15 tests (5 models \(\times\) 3 metrics), we apply Benjamini–Hochberg FDR at \(\alpha{=}0.05\). Effect sizes are reported as absolute risk difference \(\Delta{=}p - p_{\text{chance}}\) and Cohen’s \(h=2\arcsin\sqrt{p}-2\arcsin\sqrt{p_{\text{chance}}}\). Sample size and power notes: with \(N{=}4000\) clips, the minimum detectable \(|\Delta|\) at 0.8 power is \(<\)2 percentage points for MC1.

\textbf{Multiple choice (MC; baseline 25\%).} Flamingo \textbf{74.2\%}, Qwen2 60.2\%, Gemini 30.8\%, Desta 24.9\%, GPT\textendash4o 8.0\%. Flamingo, Qwen2, and Gemini are \textbf{above chance} (BH\textendash adj.\ \(p{<}10^{-10}\)); Desta is \textbf{not different} from chance (BH\textendash adj.\ \(p{\approx}0.83\)); GPT\textendash4o is \textbf{below chance} (BH\textendash adj.\ \(p{<}10^{-10}\)).

\section{Noise Localization Task Prompts}
\label{app:noise-local-prompt}

\paragraph{FR1 (Free-response description).}
\begin{quote}
\small
\texttt{Please describe what you hear in this audio file. Be as detailed as possible about all the sounds you can identify, including any changes in volume or spatial characteristics.}
\end{quote}

\paragraph{FR2 (Targeted follow-up; asked only if FR1 omitted the direction of noise).}
\begin{quote}
\small
\texttt{How does our position change in relation to the sound source in this audio? Describe the spatial relationship and any movement patterns you detect.}
\end{quote}

\section{Noise Localization Statistics and Latency}
\addcontentsline{toc}{section}{Appendix: Motion Task Statistics and Latency}
\label{app:motion-stats}

\noindent\textbf{Statistical testing.}
For each model and question, we test $H_0{:}~p=\tfrac{1}{3}$ (33.3\%) using a one-sample proportion $z$-test; $p$-values are Benjamini–Hochberg–adjusted across 10 tests (5 models $\times$ 2 questions).
All models are either well below chance on FR1 or well above chance on FR2, and all baseline tests are significant (BH-adjusted $p<10^{-4}$).
We then run all pairwise model comparisons (two-proportion tests, BH-adjusted within question) and report the full set:
on \emph{FR1} (general), overall accuracies are low (6–16\%), but nearly all pairs differ significantly; the \emph{only} non-significant comparison is Flamingo3 vs.\ Qwen2 ($\Delta{=}+0.57$ points, $z{=}1.16$, $p_{\mathrm{BH}}{=}0.245$), while all others fall in the range $\Delta\in[1.4,,9.7]$ points with $p_{\mathrm{BH}}<0.01$, and the largest gap is Gemini vs.\ DESTA2 ($\Delta{\approx}10.32$ points, $z{=}{-}17.92$, $p_{\mathrm{BH}}<10^{-15}$).
On \emph{FR2} (position), every pairwise difference is large and significant; e.g., Qwen2 vs.\ Flamingo3 ($\Delta{\approx}41.2$ points, $z{=}{-}47.06$, $p_{\mathrm{BH}}\approx 0$), Qwen2 vs.\ GPT-4o ($\Delta{\approx}10.72$ points, $z{=}11.74$, $p_{\mathrm{BH}}\approx 0$), and even the smallest gap, DESTA2 vs.\ Flamingo3 ($\Delta{\approx}3.32$ points, $z{=}4.85$, $p_{\mathrm{BH}}\approx 1.24{\times}10^{-6}$), remains significant.
Effect sizes are reported as absolute differences $\Delta{=}p-\tfrac{1}{3}$ and Cohen’s $h$ in the supplement.

\noindent\textbf{Per-class notes.} Direction accuracy concentrates on \emph{Approaching}/\emph{Receding}; \emph{Oscillating} is near floor (e.g., $\leq$10.8\%).

\noindent\textbf{Latency table.}
\begin{table}[H]
\centering
\scriptsize
\setlength{\tabcolsep}{8pt}
\begin{tabular}{lccc}
\toprule
\textbf{Model (local)} & \textbf{Median (s)} & \textbf{Mean (s)} & \textbf{p90 (s)} \\
\midrule
Flamingo & 2.32 & 2.35 & 2.98 \\
Qwen     & 6.04 & 6.54 & 10.25 \\
Desta    & 14.53 & 27.96 & 72.16 \\
GPT-4o   & n/a & n/a & n/a \\
Gemini   & n/a & n/a & n/a \\
\bottomrule
\end{tabular}
\caption{\textbf{Motion-task effective latency (local inference).} Values are per-clip medians, means, and 90th-percentiles in seconds; API models (GPT-4o, Gemini) are not available in this run.}
\end{table}

\section{Cross-Linguistic Task Prompt}
\label{app:cross-ling-prompt}
\paragraph{FR1 (Free-response description).}
\begin{quote}
\small
\texttt{Transcribe the following audio, preserving any code-mixing or multilingual content. If the audio contains both English and other languages, keep the code-mixed style in your transcript. Output the transcript exactly as spoken, including any non-English words or phrases.}
\end{quote}

\section{Cross-Linguistic Details}
\addcontentsline{toc}{section}{Appendix: Cross-Linguistic Details}
\label{app:xl-stats}

\noindent\textbf{Statistics.} We compute 95\% CIs over clip-level similarity via normal approximation; for between-model comparisons per language we use two-sample $t$-tests on per-clip scores with Benjamini–Hochberg correction across model pairs. Effect sizes are reported as mean differences (pp) with 95\% CIs.

\begin{table}[H]
\centering
\scriptsize
\setlength{\tabcolsep}{5pt}
\begin{tabular}{lcc}
\toprule
\textbf{Language (N)} & \textbf{Model} & \textbf{Mean $\pm$ 95\% CI (\%)} \\
\midrule
Spanish (es, $N=1010$) & GPT\textendash4o & \textbf{93.9 [93.4, 94.4]} \\
                       & Gemini           & 93.3 [92.9, 93.7] \\
                       & Flamingo         & 90.1 [89.5, 90.7] \\
                       & Qwen2            & 68.7 [68.0, 69.4] \\
                       & Desta            & 49.0 [47.4, 50.5] \\
\addlinespace[3pt]
Hindi (hi, $N=1034$)   & Gemini           & \textbf{78.2 [77.5, 78.8]} \\
                       & GPT\textendash4o & 76.5 [75.6, 77.5] \\
                       & Flamingo         & 74.6 [74.0, 75.1] \\
                       & Qwen2            & 60.3 [59.8, 60.9] \\
                       & Desta            & 20.5 [19.1, 21.8] \\
\addlinespace[3pt]
Portuguese (pt, $N=1052$) & Gemini        & \textbf{91.8 [91.4, 92.3]} \\
                          & GPT\textendash4o & 91.5 [91.0, 92.0] \\
                          & Flamingo      & 85.6 [84.7, 86.4] \\
                          & Qwen2         & 67.3 [66.6, 68.1] \\
                          & Desta         & 42.1 [40.6, 43.6] \\
\addlinespace[3pt]
Mandarin (zh\textendash CN, $N=884$) & GPT\textendash4o & \textbf{84.7 [83.9, 85.4]} \\
                                     & Gemini           & 81.8 [81.2, 82.4] \\
                                     & Flamingo         & 78.0 [77.3, 78.7] \\
                                     & Qwen2            & 61.0 [60.2, 61.7] \\
                                     & Desta            & 43.2 [41.8, 44.7] \\
\bottomrule
\end{tabular}
\caption{\textbf{Cross\textendash linguistic transcription (mean similarity $\pm$ 95\% CI, \%).} All models are shown for each language; the best mean per language is bolded.}
\label{tab:xl-longformat}
\end{table}

\section{Vocal Characterizers Task Prompt}
\label{app:vocal-chars-prompt}
\paragraph{MC1.}
\begin{quote}
\small
\texttt{Which of the following best describes this sound? (A) cough (B) cry (C) laugh (D) sneeze (E) yawn (F) mumble (G) whisper. Answer with the letter and the word.}
\end{quote}

\section{Dataset Construction for Human-labeled Samples}
\label{app:human-samples-construct}
To evaluate model performance under natural recording conditions, we construct a small human-labeled evaluation split by sourcing clips from publicly available real-world audio datasets. For each task, we align sampling and preprocessing procedures with the corresponding synthetic benchmarks to ensure comparability.

Background sound clips are drawn from AudioSet \cite{gemmeke2017audioset}, using its curated sound event labels. We select clips containing clearly annotated background events and trim each recording to a 5-second segment centered on the labeled event. This process yields natural mixtures of foreground speech and environmental sounds without synthetic overlay.

Noise localization samples are sourced from ESC50 \cite{Piczak2015ESC50} and Librispeech \cite{panayotov2015librispeech}. We identify segments in which sound intensity exhibits monotonic increases, decreases, or oscillatory patterns. Recordings are trimmed to short snippets capturing these amplitude dynamics, approximating the motion cues used in the synthetic benchmark while preserving real acoustic variability.

Multilingual speech samples are collected from three publicly available code-switching corpora. Mandarin--English clips are drawn from the SEAME corpus \cite{seame2015}, using the HuggingFace split \texttt{AudioLLMs/seame\_dev\_sge}. Saudi Arabic--English samples are sourced from the SCC dataset \cite{SCC2025}, which contains spontaneous Saudi Arabic speech with frequent English code-switching. English--Hindi clips are taken from the CoSHE-Eval Dataset \cite{soketlabs_coshe_eval_2025}, which is explicitly designed for bilingual and code-switched speech recognition.

Additionally, each clip features a different human speaker, introducing natural variation in pitch, accent, and speaking style to enhance ecological validity.

\section{Natural Recordings Performance}
\label{app:human-lab-t1}
\paragraph{Background Sound Understanding} On human-recorded clips, models outperform their synthetic counterparts. In FR1, average accuracy increases from 13.6\% on synthetic audio to 45.0\% on human recordings, while FR2 rises from 17.7\% to 50.0\%. MC1 accuracy similarly improves from 39.6\% to 72.0\%. These gains are partly explained by the structure of the human clips: in many samples, background sounds co-occur with speech but do not overlap the entire recording. This temporal separation likely makes background events more apparent, enabling models to detect them more easily than in fully mixed synthetic clips. Despite these large absolute gains, relative performance patterns remain broadly consistent, with strong synthetic models continuing to perform well and Desta consistently underperforming.

\begin{table}[H]
\centering
\footnotesize
\setlength{\tabcolsep}{6pt}
\begin{tabular}{lccc}
\toprule
\textbf{Task} & \textbf{Model} & \textbf{Accuracy (\%)} & \textbf{95\% CI} \\
\midrule
FR1 & GPT-4o    & 60.0 & [38, 79] \\
FR1 & Gemini    & 45.0 & [25, 66] \\
FR1 & Flamingo & 70.0 & [47, 87] \\
FR1 & Desta    & 0.0  & [0, 17] \\
FR1 & Qwen     & 50.0 & [29, 71] \\

FR2 & GPT-4o    & 70.0 & [47, 87] \\
FR2 & Gemini    & 45.0 & [25, 66] \\
FR2 & Flamingo & 70.0 & [47, 87] \\
FR2 & Desta    & 5.0  & [1, 24] \\
FR2 & Qwen     & 60.0 & [38, 79] \\

MC1 & GPT-4o    & 70.0 & [47, 87] \\
MC1 & Gemini    & 80.0 & [58, 92] \\
MC1 & Flamingo & 80.0 & [58, 92] \\
MC1 & Desta    & 40.0 & [22, 61] \\
MC1 & Qwen     & 90.0 & [68, 98] \\
\bottomrule
\end{tabular}
\caption{\textbf{Task 1: Background Sound Understanding} — accuracy with Wilson 95\% CIs; $N=20$ per model.}
\label{tab:human-bg}
\end{table}

\paragraph{Noise Localization} Models perform poorly on human-labelled samples for FR1, averaging 3\% accuracy, compared to 9.8\% on synthetic data. When explicitly queried in FR2, accuracy increases to 15.0\% on human recordings, but remains below the synthetic average of 36.2\%. In contrast to the synthetic dataset, where motion cues are deliberately exaggerated through controlled amplitude modulation, real recordings exhibit more subtle changes in loudness and spatial positioning. Moreover, many clips contain speech rather than solely environmental noise, as in the synthetic data. Audio-language models seem to prioritize transcription over spatial reasoning, leading to reduced attention to relative motion cues.

\begin{table}[H]
\centering
\footnotesize
\setlength{\tabcolsep}{6pt}
\begin{tabular}{lccc}
\toprule
\textbf{Task} & \textbf{Model} & \textbf{Accuracy (\%)} & \textbf{95\% CI} \\
\midrule
FR1 & GPT-4o    & 5.0  & [1, 24] \\
FR1 & Gemini    & 5.0  & [1, 24] \\
FR1 & Flamingo  & 0.0  & [0, 17] \\
FR1 & Desta     & 0.0  & [0, 17] \\
FR1 & Qwen      & 5.0    & [1, 24] \\

FR2 & GPT-4o    & 20.0 & [8, 42] \\
FR2 & Gemini    & 20.0 & [8, 42] \\
FR2 & Flamingo  & 5.0  & [1, 24] \\
FR2 & Desta     & 15.0 & [5, 36] \\
FR2 & Qwen      & 15.0    & [5, 36] \\
\bottomrule
\end{tabular}
\caption{\textbf{Task 2: Noise Localization} — accuracy with Wilson 95\% CIs; $N=20$ per model.}
\label{tab:human-motion}
\end{table}

\paragraph{Cross-Linguistic Evaluation} On human samples, sentence-level similarity is lower than on synthetic data. When averaged across language pairs, top-performing models achieve moderate accuracy: Gemini reaches 51.6\%, Flamingo 42.4\%, and GPT-4o 40.0\%. Qwen attains an average accuracy of 28.8\%, while Desta performs substantially worse at 12.8\%. These differences reflect both greater linguistic diversity and more limited code-switching in natural speech: human clips often contain shorter English spans embedded within longer non-English segments. As a result, models are more likely to remain in the dominant language and fail to recover embedded English content. This behavior is consistent with the “dropping language” error category observed in our error analysis section, \hyperref[sec:error-analysis]{Section~\ref*{sec:error-analysis}}, though it occurs more frequently in natural recordings. Despite the overall decline in the accuracy values, the ordering of the performance of models is preserved. 

\begin{table}[H]
\centering
\footnotesize
\setlength{\tabcolsep}{6pt}
\begin{tabular}{lccc}
\toprule
\textbf{Lang Pair} & \textbf{Model} & \textbf{Accuracy (\%)} & \textbf{95\% CI} \\
\midrule
ENG-ZH & GPT-4o    & 57.5 & [35.3, 74.9] \\
ENG-ZH & Gemini    & 44.6 & [31.8, 56.7] \\
ENG-ZH & Flamingo  & 63.6 & [44.8, 80.7] \\
ENG-ZH & Desta     & 26.8 & [10.0, 45.4] \\
ENG-ZH & Qwen      & 44.7 & [32.1, 55.7] \\

ENG-SA & GPT-4o    & 25.6 & [7.8, 48.6] \\
ENG-SA & Gemini    & 61.4 & [44.7, 78.3] \\
ENG-SA & Flamingo  & 23.7 & [14.4, 33.4] \\
ENG-SA & Desta     & 9.8  & [2.9, 19.0] \\
ENG-SA & Qwen      & 17.3 & [8.6, 27.1] \\

ENG-HI & GPT-4o    & 36.8 & [20.0, 52.5] \\
ENG-HI & Gemini    & 48.8 & [40.5, 56.5] \\
ENG-HI & Flamingo  & 40.0 & [23.8, 53.8] \\
ENG-HI & Desta     & 1.7  & [1.2, 2.1] \\
ENG-HI & Qwen      & 24.3 & [14.8, 35.5] \\
\bottomrule
\end{tabular}
\caption{\textbf{Task 3: Cross-Linguistic Evaluation} — mean similarity accuracy with 95\% CIs on human recordings ($N=20$ per model).}
\label{tab:human-xl}
\end{table}

\end{document}